\documentclass[10pt, conference, compsocconf]{IEEEtran}
\ifCLASSINFOpdf
\else
\fi
\usepackage{url}

\usepackage{graphicx}
\usepackage[caption=false]{subfig}
\usepackage{multirow}
\usepackage{array}
\usepackage{mdwlist} 
\usepackage{flushend}

\begin{document}
%
\title{Boosting Metrics for Cloud Services Evaluation --- The Last Mile of Using Benchmark Suites}


\author{\IEEEauthorblockN{Zheng Li}
\IEEEauthorblockA{School of CS\\
NICTA and ANU\\
Canberra, Australia\\
Zheng.Li@nicta.com.au}
\and
\IEEEauthorblockN{Liam O'Brien}
\IEEEauthorblockA{Research School of CS\\
CECS, ANU\\
Canberra, Australia\\
liamob99@hotmail.com}
\and
\IEEEauthorblockN{He Zhang}
\IEEEauthorblockA{School of CSE\\
NICTA and UNSW\\
Sydney, Australia\\
He.Zhang@nicta.com.au}
\and
\IEEEauthorblockN{Rainbow Cai}
\IEEEauthorblockA{Division of Information\\
Information Services, ANU\\
Canberra, Australia\\
Rainbow.Cai@anu.edu.au}
}


%


\maketitle

\begin{abstract}
Benchmark suites are significant for evaluating various aspects of Cloud services from a holistic view. However, there is still a gap between using benchmark suites and achieving holistic impression of the evaluated Cloud services. Most Cloud service evaluation work intended to report individual benchmarking results without delivering summary measures. As a result, it could be still hard for customers with such evaluation reports to understand an evaluated Cloud service from a global perspective. Inspired by the boosting approaches to machine learning, we proposed the concept \textit{Boosting Metrics} to represent all the potential approaches that are able to integrate a suite of benchmarking results. This paper introduces two types of preliminary boosting metrics, and demonstrates how the boosting metrics can be used to supplement primary measures of individual Cloud service features. In particular, boosting metrics can play a summary \textit{Response} role in applying experimental design to Cloud services evaluation. Although the concept \textit{Boosting Metrics} was refined based on our work in the Cloud Computing domain, we believe it can be easily adapted to the evaluation work of other computing paradigms. 

\end{abstract}


%
\IEEEpeerreviewmaketitle

\section{Introduction}
As Cloud Computing becomes one of the most promising computing paradigms in industry \cite{Buyya_Yeo_2009}, numerous vendors have started to supply public Cloud infrastructures and services \cite{Prodan_Ostermann_2009}. Since most providers do not reveal details about their infrastructures \cite{Brooks_2010}, customers have little knowledge and control
over the precise nature of public Cloud services even in the ``locked down" environment \cite{Sobel_Subramanyam_2008}. As such, Cloud services evaluation would be crucial and beneficial for both service customers (e.g. cost-benefit analysis) and providers (e.g. direction of improvement) \cite{Li_Yang_2010}. 

When it comes to evaluating a Cloud service, a benchmark suite is usually required to cover and test various aspects of the service from a holistic view \cite{Iosup_Ostermann_2011,Rabl_Frank_2010}. In practice, however, there is still a gap between using benchmark suites and achieving such holistic impressions. Through reviewing the existing studies of Cloud services evaluation \cite{Li_Zhang}, we found that, even if benchmark suites were adopted to evaluate Cloud services, most evaluators intended to report individual benchmarking results with a lack of visibly integrated measurements. As a result, customers with such evaluation reports could have to further summarize various evaluation results by themselves. Although it is possible and sometimes flexible for customers to balance tradeoffs in employing a Cloud service, it is often useful and convenient to compare alternatives by using a single index \cite{Islam_Lee_2012}, and such a single index can act as a supplementary to individual benchmarking results for customers' decision making. More importantly, a single measurement that reflects the overall performance of a Cloud service can play a summary \textit{Response} role in experimental design and analysis \cite{Montgomery_2009} for evaluating the Cloud service. Therefore, to facilitate measuring Cloud service performance as a whole, it is valuable and necessary to investigate approaches to integration of a suite of benchmarking results. 

Inspired by the boosting approaches to machine learning that combine weak rules into a single more accurate one \cite{Schapire_2002}, we proposed the concept \textit{Boosting Metrics} to represent all the potential approaches that are able to integrate a suite of benchmarking results. A boosting metric can then be viewed as a secondary measure by manipulating the primary metrics that directly measure individual Cloud service aspects. To clearly demonstrate this proposed concept, this paper rationalizes several common and straightforward approaches to boosting metrics instead of demonstrating sophisticated ones. Moreover, a real case of investigating global performance of two Amazon EC2 instances is studied to show how the boosting metrics can help facilitate the corresponding experimental design and analysis. Note that, although the concept \textit{Boosting Metrics} was refined based on our work in the Cloud Computing domain, we believe it can be easily adapted to the evaluation work of other computing paradigms. 

The remainder of this paper is organized as follows. The existing Cloud services evaluation work related to \textit{Boosting Metrics} are briefly summarized in Section \ref{IV}. Section \ref{II} specifies the proposed concept \textit{Boosting Metrics}, and then introduces two types of preliminary approaches to boosting metrics. Section \ref{III} employs a case study of evaluating Amazon EC2 to demonstrate how a boosting metric can be used in evaluation and how it can help analyze experimental results. Conclusions and some future work are discussed in Section \ref{V}.



\section{Related Work}
\label{IV}
According to the review on evaluating commercial Cloud services \cite{Li_Zhang}, most of the existing studies employed benchmark suites or multiple benchmarks to evaluate Cloud services, while only several practices directly showed overall measurements by using boosting metrics. 

\subsection{Usage of Benchmark suites in Cloud Services Evaluation}
Since one Cloud service may have various features and aspects, it has been identified that benchmark suites are usually required for Cloud services evaluation to cover and test various service aspects from a holistic view \cite{Iosup_Ostermann_2011,Rabl_Frank_2010}. In fact, different benchmark suites have already been widely employed in the existing practices of Cloud services evaluation. For example, the kernel benchmarks in NPB have been used to reveal different micro features of Amazon EC2 like computation, communication and storage respectively \cite{Akioka_Muraoka_2010}; while six scale-out workloads are collected to simulate different macro application scenarios in today's Cloud infrastructure \cite{Ferdman_Adileh_2012}. In particular, for verifying scientific computing in the Cloud, HPCC seems a popular benchmark suite to show high performance computing capabilities of Cloud services\cite{Jackson_Ramakrishnan_2010,Ostermann_Iosup_2009}. In addition to those predefined benchmark suites, the evaluator-selected application sets were also commonly adopted to evaluate Cloud services \cite{Dejun_Pierre_2009,Jackson_Ramakrishnan_2010}. Essentially, each application set here can be viewed as an individual benchmark suite.

\subsection{Usage of Boosting Metrics in Cloud Services Evaluation}
\label{relatedMetrics}
Although not common, the idea of boosting metrics has been intuitively employed in some Cloud services evaluation work, together with a little preliminary discussion about the merits of employing boosting metrics.
For example, the geometric mean of eight NAS Parallel Benchmarks (NPB) results (BT, CG, FT, IS, LU, MG, SP, UA) was used to measure the computational performance of Amazon EC2 on a wide set of model applications and kernels \cite{Evangelinos_Hill_2008}. 
A more sophisticated sample is the usage of metric \textit{Sustained System Performance (SSP)}, which combined a set of application measurements to give an aggregate measure of overall performance of a Cloud service \cite{Jackson_Ramakrishnan_2010}. In detail, the calculation of SSP is to multiply the geometric mean of individual applications' performance per CPU core by the number of computational cores, which can also be viewed as a \textit{Geometric Mean}-based boosting metric with respect to the application benchmark suite.

\section{Preliminary Approaches to Boosting Metrics}
\label{II}
As mentioned previously, we borrowed the ``boosting" idea from the machine learning field to our Cloud services evaluation work. In machine learning, boosting refers to the method of producing a more accurate prediction rule by combining a set of rough and less accurate rules of thumb \cite{Schapire_2002}. By analogy, in Cloud services evaluation, we treat ``boosting" as integrating a suite of local benchmarking results into a single global measurement, namely boosting metrics, to reflect the overall performance of a Cloud service. In particular, boosting metrics can be regarded as secondary measures based on the primary metrics that measure individual Cloud service aspects. When measuring different Cloud service aspects, benchmark suites may adopt homogeneous primary metrics (e.g.~NPB \cite{NASA_2012}) or inhomogeneous primary metrics (e.g.~HPCC \cite{Luszczek_Bailey_2006}). Correspondingly, here we show two types of boosting metrics, as described in the following two subsections respectively.

\subsection{Mean as a Boosting Metric from a Spatial Perspective}
\begin{figure}[!t]
\centering
\includegraphics[width=6.5cm]{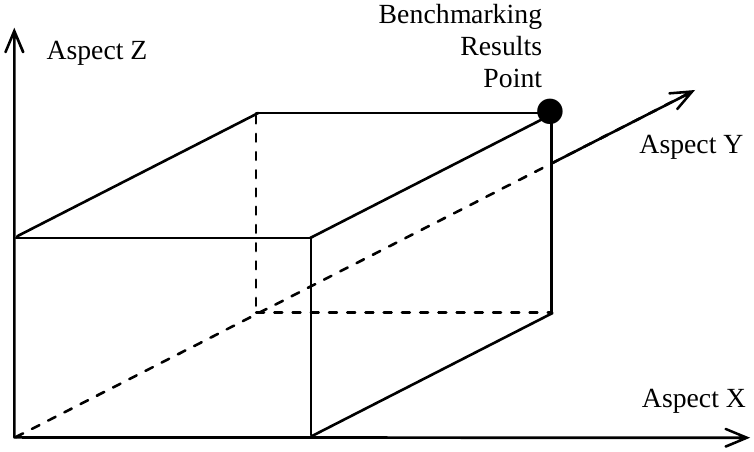}
\caption{\label{fig>PicPoint}The benchmarking results point in a Cloud service aspect space.}
\end{figure}

In a benchmark suite, a set of different benchmarks are generally expected to be able to reflect different aspects of a Cloud service in an evaluation. If we imagine every single service aspect as an individual dimension, a Cloud service with \textit{n} aspects can be represented as a Euclidean \textit{n}-space. As such, when evaluating a Cloud service, the benchmarking results together would identify a particular point in the Euclidean \textit{n}-space, which essentially uses a tuple to reflect the overall feature of the Cloud service with respect to the corresponding benchmark suite, as illustrated in Figure \ref{fig>PicPoint}. 

As previously mentioned, a boosting metric is supposed to represent overall service feature by using a single number instead of using a tuple. Thus, seeking boosting metrics here is to find single-number representations of the benchmarking point in the service-aspect space. Since the benchmarking point and the origin can determine a rectangular parallelepiped in the Euclidean \textit{n}-space (cf.~Figure \ref{fig>PicPoint}), we can switch our focus from the coordinates of the point to the attributes of the corresponding rectangular parallelepiped, such as the perimeter, surface area, volume, etc. Furthermore, given the related work, we tried to rationalize several ``classic" means \cite{Cantrell_2012} to suit the rectangular parallelepiped's attributes, rather than re-inventing ``new" measures. The equations of the selected means are listed below, where $Benchmarking_i$ denotes the benchmarking result by using the \textit{i}th benchmark in a suite. 

\subsubsection{Arithmetic Mean}
Corresponding to that the perimeter of a rectangular parallelepiped is the sum of its side lengths, we may use the arithmetic mean as a potential boosting metric, as shown in Equation (\ref{equation>1}).

\begin{small}
\begin{equation}
\label{equation>1}
A=\frac{\sum\limits_{i=1}^n Benchmarking_i}{n}
\end{equation}
\end{small}

\subsubsection{Geometric Mean}
Corresponding to that the volume of a rectangular parallelepiped is the product of its side lengths, we may use the geometric mean as a candidate boosting metric, as shown in Equation (\ref{equation>2}).

\begin{small}
\begin{equation}
\label{equation>2}
G=\sqrt[n]{\prod_{i=1}^n Benchmarking_i} 
\end{equation}
\end{small}

\subsubsection{Harmonic Mean}
In particular, corresponding to the rate between the volume and surface area of a rectangular parallelepiped, we may use the harmonic mean as a candidate boosting metric, as shown in Equation (\ref{equation>3}).

\begin{small}
\begin{equation}
\label{equation>3}
H=\frac{n\times\prod\limits_{j=1}^n Benchmarking_j}{\sum\limits_{i=1}^n \frac{\prod\limits_{j=1}^n Benchmarking_j}{Benchmarking_i}}=\frac{n}{\sum\limits_{i=1}^n \frac{1}{Benchmarking_i}}
\end{equation}
\end{small}

\subsubsection{Quadratic Mean}
Corresponding to the distance between the benchmarking point and the origin, we may use the quadratic mean as a candidate boosting metric, as shown in Equation (\ref{equation>4}).

\begin{small}
\begin{equation}
\label{equation>4}
Q=\sqrt{\frac{\sum\limits_{i=1}^n Benchmarking_i^2}{n}}
\end{equation}
\end{small}

As can be seen, it is convenient to calculate these means of a set of benchmarking results to reflect the summary feature of a Cloud service. Interestingly, the \textit{Geometric Mean} seems the most popular one in practice \cite{Evangelinos_Hill_2008,Jackson_Ramakrishnan_2010}. Nevertheless, there is a default constraint when employing means as boosting metrics: different Cloud service aspects should be homogeneously measured by using different benchmarks in a suite. In other words, to calculate means (secondary metrics), different benchmarking results for different Cloud service aspects must adopt the same primary metric. If the constraint cannot be satisfied, we may employ a more generic solution -- \textit{Radar Plot}, as specified in the following subsection.

\subsection{Radar Plot as a Boosting Metric}
\begin{figure}[!t]
\centering
\includegraphics[width=3.2in]{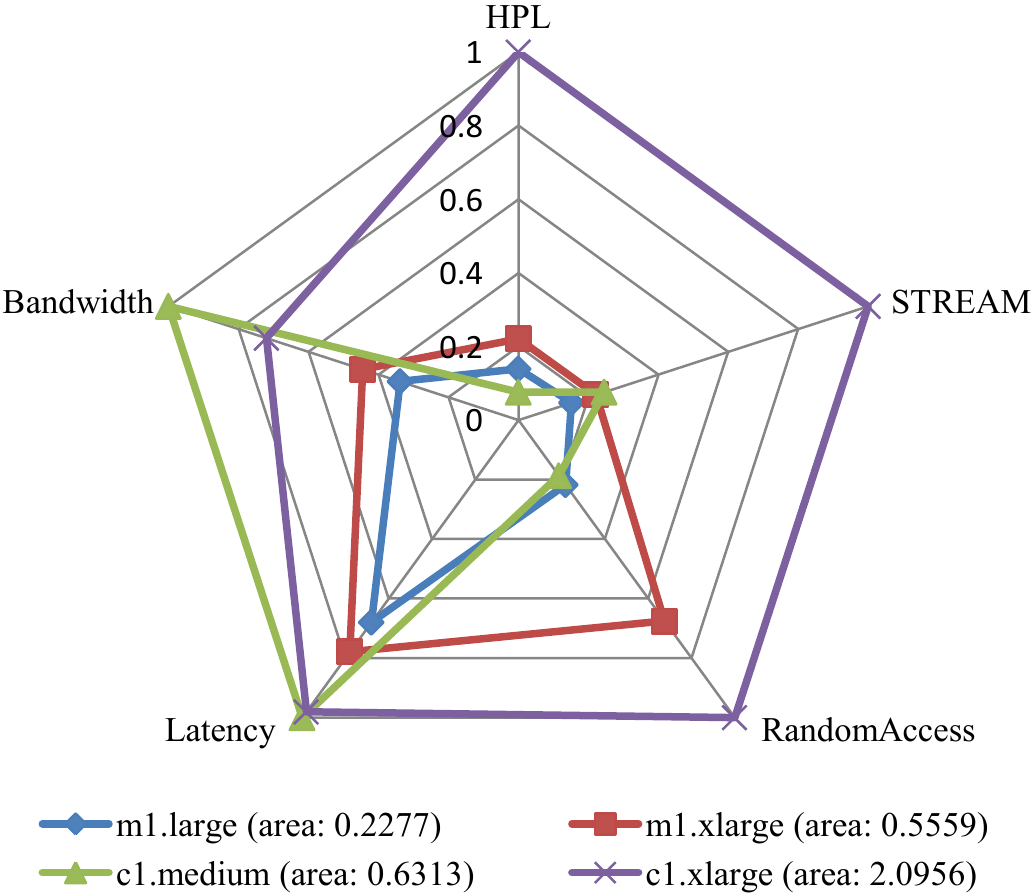}
\caption{\label{fig>PicRadar}The benchmarking results point in a Cloud service aspect space.}
\end{figure}

Radar plot is a simple but intuitive graphical tool that can simultaneously depict a group of different types of values relative to a central point. When a benchmark suite uses different primary metrics to measure different Cloud service aspects, we can use radar plot to represent the benchmarking results over a predefined baseline. In particular, we can also portray several groups of standardized benchmarking results in one radar plot without predefining any baseline (cf.~Figure \ref{fig>PicRadar}). Given the analysis of the existing metrics for Cloud services evaluation \cite{Li_OBrien_2012b}, here we elaborate two standardization methods only for Higher Better (HB) metrics and Lower Better (LB) metrics \cite{Obaidat_Boudriga_2010} respectively.

\begin{small}
\begin{equation}
\label{equation>5}
HB\_Standardized_i=\frac{Benchmarking_i}{MAX(Benchmarking_{1, ..., n})}
\end{equation}
\end{small}

\begin{small}
\begin{equation}
\label{equation>6}
LB\_Standardized_i=\frac{\frac{1}{Benchmarking_i}}{MAX(\frac{1}{Benchmarking_{1, ..., n}})}
\end{equation}
\end{small}

Equation (\ref{equation>5}) is for the standardization of HB metrics, while Equation (\ref{equation>6}) for LB metrics. Here $Standardized_i$ refers to the stardardized \textit{i}th benchmarking result $Benchmarking_i$. In fact, Equation (\ref{equation>6}) offers LB metrics a higher better representation through reciprocal standardization, so that all the standardized benchmarking results can be settled homogeneously higher better in a radar plot, and meanwhile construct a bigger-area better polygon. As such, we can intuitively contrast the areas of different polygons to compare different groups of benchmarking results. Moreover, the area of a polygon can be regarded as a single numerical \textit{Response} to facilitate experimental design and analysis. Suppose there are \textit{n} benchmarking results standardized and marked in a radar plot, we can calculate the area of the corresponding polygon by summing up areas of the \textit{n} adjacent triangles, as shown in Equation \ref{equation>7}.

\begin{small}
\begin{equation}
\label{equation>7}
\sum\limits_{i=1}^n \frac{\sin(\frac{2\pi}{n})\times Standardized_i\times Standardized_{mod(i+1,n)}}{2}
\end{equation}
\end{small}

Here we employ a real case to demonstrate the radar plot as a boosting metric. For our convenience, the evaluation data reported in \cite{Ostermann_Iosup_2009} are directly reused, as shown in Table \ref{table>original}. Given the various types of benchmarking results, such as HPL, STREAM, RandomAccess, Latency, and Bandwidth, within the HPCC benchmark suite \cite{Luszczek_Bailey_2006}, it is hard to compare the summary performance as a whole when evaluating different types of EC2 instances. 

\begin{table}[!t]\footnotesize
\renewcommand{\arraystretch}{1.3}
\centering
\caption{\label{table>original}Original HPCC Benchmarking Results for Various EC2 Instance Types}
\begin{tabular}{|m{1.85cm}|m{1cm}|m{1.1cm}|m{1.2cm}|m{1cm}|m{1cm}|}
\hline
\textbf{Name} & \textbf{m1.large} & \textbf{m1.xlarge} & \textbf{c1.medium} & \textbf{c1.xlarge}\\
\hline
HPL (GFLOPS)& 7.15 & 11.38 & 3.91& 51.58\\
\hline
STREAM (GBps) & 2.38 & 3.47 & 3.84 & 15.65\\
\hline
RandomAccess (MUPs) & 54.35 & 168.64 & 46.73 & 249.66\\
\hline
Latency ($\mu$s) & 20.48 & 17.87 & 13.92 & 14.19\\
\hline
Bandwidth (GBps) & 0.7 & 0.92 & 2.07 & 1.49\\
\hline
\end{tabular}
\end{table}

Thus, we first standardize those benchmarking results respectively, as listed in Table \ref{table>standard}. Note that the generated numbers in Table \ref{table>standard} do not come with any benchmarking unit. Then, the standardized benchmarking results are represented in a radar plot, as illustrated in Figure \ref{fig>PicRadar}. Through this radar plot, we can intuitively and conveniently identify that: c1.xlarge has absolutely better overall performance than m1.large and m1.xlarge; c1.xlarge is also better than c1.medium in general, while slightly poorer in terms of Bandwidth and Latency. In particular, the areas of different benchmarking polygons in the radar plot are further calculated to quantitatively reflect the summary performance of the four types of EC2 instances, as bracketed beside the legend entries. Essentially, the numerical areas may play a \textit{Response} role in the design of experiments for Cloud services evaluation. A complete experimental design and analysis sample by using boosting metrics is elaborated in the next section.

\begin{table}[!t]\footnotesize
\renewcommand{\arraystretch}{1.3}
\centering
\caption{\label{table>standard}Standardized HPCC Benchmarking Results for Various EC2 Instance Types}
\begin{tabular}{|m{1.85cm}|m{1cm}|m{1.1cm}|m{1.2cm}|m{1cm}|m{1cm}|}
\hline
\textbf{Name} & \textbf{m1.large} & \textbf{m1.xlarge} & \textbf{c1.medium} & \textbf{c1.xlarge}\\
\hline
HPL & 0.1386 & 0.2206 & 0.0758 & 1\\
\hline
STREAM & 0.1521 & 0.2217 & 0.2454 & 1\\
\hline
RandomAccess & 0.2177 & 0.6755 & 0.1872 & 1\\
\hline
Latency & 0.6797 & 0.779 & 1 & 0.981\\
\hline
Bandwidth & 0.3382 & 0.4444 & 1 & 0.7198\\
\hline
\end{tabular}
\end{table}

\section{A Case Study of Using Boosting Metrics in Experimental Design and Analysis}
\label{III}
\subsection{Problem and Motivation}
We proposed to use a set of Amazon EC2 instances to perform a small-scale parallel computing project. According to the estimation of our project and the predefined EC2 instance types \cite{AWS_2012}, we initially selected m1.xlarge and m2.xlarge as two candidate types of parallel computing nodes. The spectifications and prices of these two EC2 instance types are listed in Table \ref{table>instance}. 
When making decision to choose the most suitable alternative from these two options, we found that it was hard to directly distinguish the better one based on their specifications. Unlike the clear differences between the other types of EC2 instances, each of these two options has its own distinctions. For example, m1.xlarge seems overall better than m2.xlarge, while m2.xlarge has faster single CPU core, larger memory, and lower price. As such, we decided to evaluate these two types of EC2 instances to roughly compare their potential performance in our project. 
It is then suitable to consider boosting metrics for summary measurement in this case.

\begin{table}[!t]\footnotesize
\renewcommand{\arraystretch}{1.3}
\centering
\caption{\label{table>instance}Specifications and Prices of Two EC2 Instance Types}
\begin{tabular}{|l|c|c|}
\hline
\textbf{Specification} & \textbf{m1.xlarge} & \textbf{m2.xlarge} \\
\hline
Core Amount & 4 & 2\\
\hline
ECU Amount & 8 & 6.5\\
\hline
Network I/O Performance & High & Moderate\\
\hline
Memory Size & 15 GB & 17.1 GB\\
\hline
Platform & 64 bit & 64 bit\\
\hline
Storage Size & 1690 GB & 420 GB\\
\hline
Windows Usage Cost & \$0.92 per Hour & \$0.57 per Hour\\
\hline
\end{tabular}
\end{table}

\subsection{Evaluation Preparation}
Given the above discussion about the evaluation requirement, a series of prerequisites can be set up as preparation for implementing evaluation experiments.

\subsubsection{Experimental Environment}
As mentioned previously, the evaluation requirement in this case can be viewed as merely a rough understanding of the parallel computing capability of those two instance types. 
To save time, we decided to perform evaluation for each option only on a single EC2 instance rather than on a real parallel cluster environment. In detail, we applied one m1.xlarge instance and one m2.xlarge instance respectively from Amazon's US-EAST-1 region, and both instances came with the same quick launch Amazon Machine Image (AMI) -- 64 bit Windows Server 2008 Base.

\subsubsection{Benchmark and Metrics Selection}  
As a well-known and well-accepted parallel computing benchmark suite, NPB has been widely used for Scientific Computing evaluation in the public Cloud \cite{Akioka_Muraoka_2010,Carlyle_Harrell_2010,Evangelinos_Hill_2008,He_Zhou_2010,Walker_2008}. Therefore, we also employed NPB as the benchmark suite to evaluate the summary performance of EC2 instances for our project. In particular, since the software system in our project was implemented using JAVA, we selected the latest JAVA version of NPB, namely NPB3.0-JAV \cite{NASA_2012}. Although different benchmarks in NPB are used to reflect different features of a computing system like computation, communication and storage, all the NPB benchmarking results adopt the same format with homogeneous metrics, such as \textit{benchmark runtime} (time in seconds) and \textit{benchmark FLOP rate} (floating point Mops total). Following the popular choice (cf.~Subsection \ref{relatedMetrics}), we also chose \textit{Geometric Mean} as the boosting metric over the primary metrics \textit{benchmark runtime} and \textit{benchmark FLOP rate} in this case study. 

\subsubsection{Experimental Factors Identification}
\label{subsub>factorIdentification}
Before evaluating a system, experimental factors identification is a tedious but necessary task \cite{leBoudec_2010}. Factors here refer to the elements in the system or the workload that may influence the evaluation result. In fact, our previous work has established a factor framework for Cloud services evaluation, and the latest framework version capsules the state-of-the-practice of performance evaluation factors that people currently take into account in the Cloud Computing domain \cite{Li_OBrien_2012c}. Since this evaluation work would also measure performance of EC2 instances, we conventionally identified experimental factors within the proposed framework. In detail, we explored experimental factors related to Cloud resource and benchmark's workload respectively: \textit{Instance Type} (m1.xlarge vs.~m2.xlarge), \textit{Thread Number} (2 vs.~4), and \textit{Workload Size} (Class W vs.~Class A) 

\subsection{Experimental Design}
When it comes to experimental design, there are three basic principles: Randomization, Replication, and Blocking \cite{Montgomery_2009}. In this case, we only focus on the Randomization and Replication. Although an entire NPB suite run can be treated as a block, here we try to simplify the demonstration without elaborating sophisticated design approaches. The detailed designing process is then composed of three steps, as specified below.

\subsubsection{Determining Individual Experimental Trials}
In this evaluation work, an experimental trial indicates a specific benchmark run on an EC2 instance. In practice, we used one batch command to drive a single NPB benchmark run during the experiments. 
Thus, a series of batch commands were listed to represent different individual experimental trials. 
\subsubsection{Determining Amount of Experimental Trials}
As mentioned previously, we decided to investigate two levels of \textit{Workload Size} (Class A and W) and two levels of \textit{Thread Number} (2 and 4) for two \textit{Instance Types} (m1.xlarge and m2.xlarge). To facilitate the investigation, we also planned benchmarking with single thread as a reference baseline. On the other hand, the JAVA-version NPB suite comprises seven benchmarks. According to our pilot test of running those seven benchmarks on a local machine, we decided to replicate all the different trials five times. Therefore, there are $2\times 3\times 7\times 5=210$ experimental trials in total on each instance.
\subsubsection{Determining Sequence of Experimental Trials} 
To achieve a randomized trial sequence, we assigned two random numbers to each trial-associated batch command in EXCEL. The 210 batch commands can be ordered by one random number and another in turn, to run experiments on the m1.xlarge and m2.xlarge instances respectively. Through such a randomization, we made individual trials as independent as possible between each other to reduce the experimental sequence-related bias.


\subsection{Experimental Result}
Due to the limit of space, the specific experimental results from individual NPB benchmarks are not reported in this paper. In summary, by averaging results of identical experimental trials, and dividing the trials into different workload-size, thread-number, and instance-type groups, we obtained a set of \textit{Runtime} and \textit{FLOP Rate} geometric means with respect to the NPB suite under different conditions, as listed in Table \ref{table>result}.

\begin{figure*}[!t]
\centering
\begin{tabular}{cc}
\subfloat[\label{subResultA}Geometric means of NPB runtime with workload Class W.]{\includegraphics[width=3in]{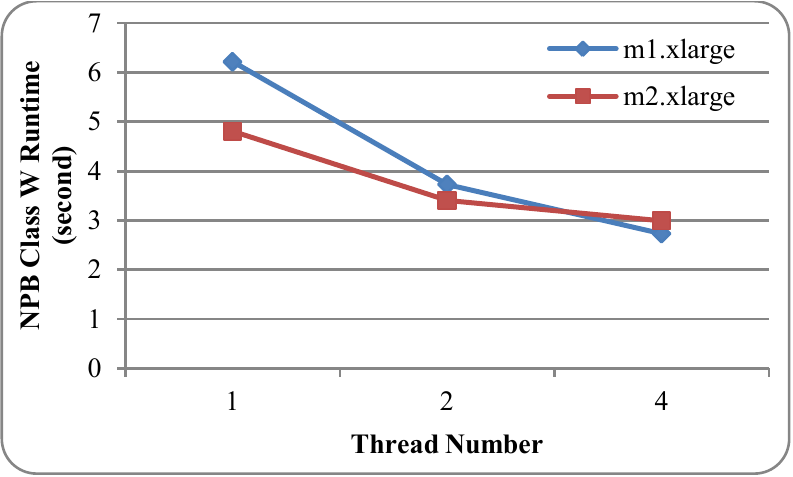}} & 
\subfloat[\label{subResultB}Geometric means of NPB runtime with workload Class A.]{\includegraphics[width=3in]{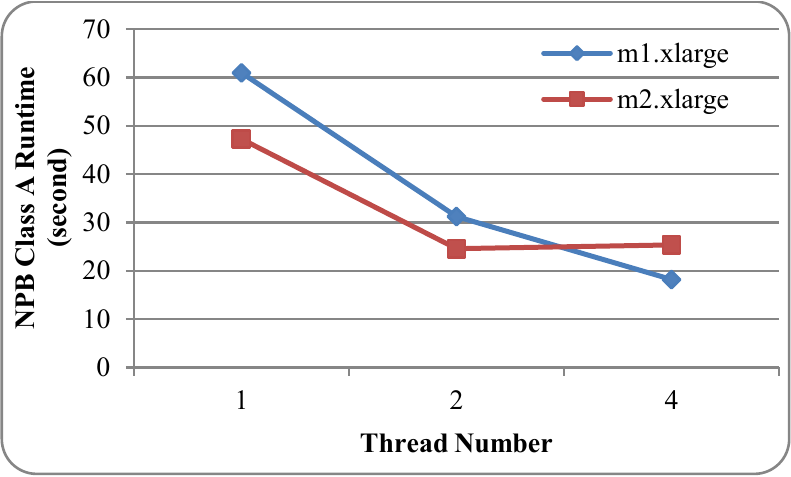}}\\
\subfloat[\label{subResultC}Geometric means of NPB FLOP rate with workload Class W.]{\includegraphics[width=3in]{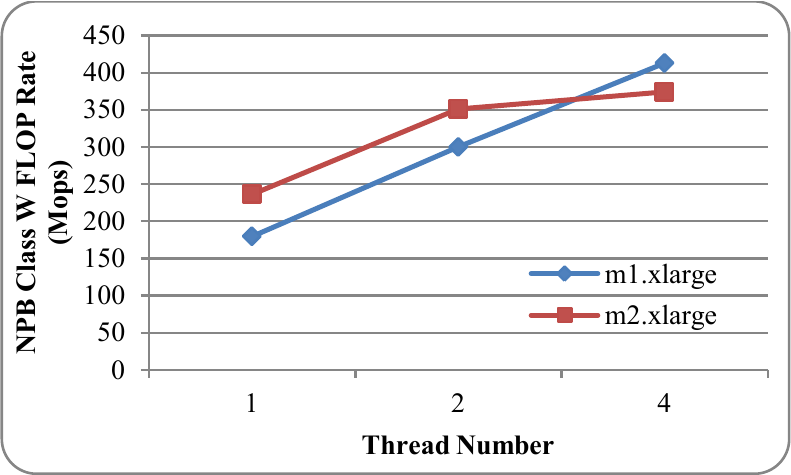}} & 
\subfloat[\label{subResultD}Geometric means of NPB FLOP rate with workload Class A.]{\includegraphics[width=3in]{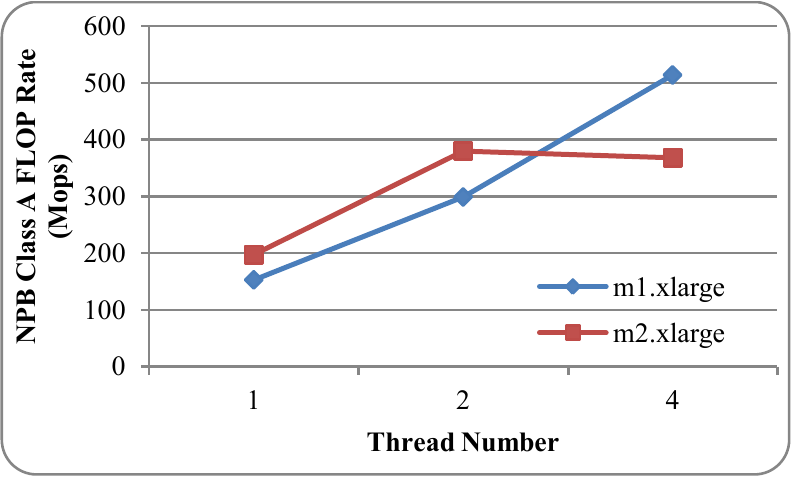}}\\
\end{tabular}
\caption{Illustration of geometric means of NPB3.0-JAV benchmarking results with different circumstances.}
\label{figure>result}
\end{figure*}

To intuitively show the instances' performance changing when varying conditions, we also used four line charts to represent the boosting metric's measurements, as illustrated in Figure \ref{figure>result}. It is not surprising that, benefiting from the faster single core, the m2.xlarge instance defeats the m1.xlarge instance for running NPB suite before over-saturating its CPU cores, while the m1.xlarge instance performs better with four-thread trials. Nevertheless, it is still hard to tell whether \textit{Instance Type} is a significant factor or not in general. Therefore, we employed formal experimental-analysis techniques to unfold further investigation, as explained in the following subsection.

\newcommand{\tabincell}[2]{\begin{tabular}{@{}#1@{}}#2\end{tabular}}
\begin{table*}[!t]\footnotesize
\renewcommand{\arraystretch}{1.3}
\centering
\caption{\label{table>result}Geometric Means of NPB3.0-JAV Benchmarking Results with Different Circumstances}
\begin{tabular}{|c|c|c|c|c|c|c|c|}
\hline
\multirow{3}{*}{\textbf{Workload}} & \textbf{\multirow{3}{*}{\tabincell{c}{Boosting Metric\\ (Geometric Mean)}}} & \multicolumn{6}{c|}{\textbf{Cloud Resource}}\\
\cline{3-8}
& & \multicolumn{3}{c|}{\textbf{EC2 Instance m1.xlarge}} & \multicolumn{3}{c|}{\textbf{EC2 Instance m2.xlarge}} \\

\cline{3-8}
& & \textbf{1 Thread} & \textbf{2 Threads} & \textbf{4 Threads} & \textbf{1 Thread} & \textbf{2 Threads} & \textbf{4 Threads}\\
\hline
\multirow{2}{*}{\textbf{Class W}} & \textbf{NPB Runtime (second)} & 6.215 & 3.727 & 2.73 & 4.808 & 3.401 & 2.987\\
\cline{2-8}
& \textbf{NPB FLOP Rate (Mops)} & 179.706 & 299.813 &412.717 & 236.363 & 351.003 & 373.948\\
\hline
\multirow{2}{*}{\textbf{Class A}} & \textbf{NPB Runtime (second)} & 60.889 & 31.176 & 18.138 & 47.221 & 24.537 & 25.32\\
\cline{2-8}
& \textbf{NPB FLOP Rate (Mops)} & 153.017 & 298.949 & 513.873 & 197.354 & 379.765 & 368.289\\
\hline
\end{tabular}
\end{table*}

\subsection{Experimental Analysis}
Since only two levels of an experimental factor were particularly concerned in this case (cf.~Subsection \ref{subsub>factorIdentification}), we naturally adopted the optimal design and analysis technique, namely Full-factorial $2^k$ Design \cite{Montgomery_2009}, to analyze the experimental results. Given the three factors considered, a pseudo $2^3$ design matrix was generated as shown in Table \ref{table>matrix}. The response columns in the matrix were filled with pseudo-trial results that correspond to eight geometric means in Table \ref{table>result}. For conciseness, we further assigned aliases to those experimental factors and responses, as listed below.

\begin{itemize*}
    \item	Factor A: Instance Type (m1.xlarge vs. m2.xlarge).
    \item	Factor B: Thread Number (2 vs. 4). 
    \item	Factor C: Workload Size (Class W vs. Class A).
    \item	Response R1: NPB Runtime (seconds).
    \item	Response R2: NPB FLOP Rate (Mops).
\end{itemize*}

\begin{table}[h]\footnotesize
\renewcommand{\arraystretch}{1.1}
\centering
\caption{\label{table>matrix}A Full-factorial ($2^3$) Design Matrix for This Case Study}
\begin{tabular}{c}
$
\left[
  \begin{array}{cccccc}
    trial & A & B & C & R1(seconds) & R2(Mops)\\
    1 & m1 & 2 & W & 3.727 & 299.813 \\
    2 & m1 & 4 & A & 18.138 & 513.873 \\
    3 & m2 & 2 & W & 3.401 & 351.003\\
    4 & m1 & 2 & A & 31.176 & 298.949\\
    5 & m2 & 2 & A & 24.537 & 379.765\\
    6 & m2 & 4 & A & 25.32 & 368.289\\
    7 & m1 & 4 & W & 2.73 & 412.717\\
    8 & m2 & 4 & W & 2.987 & 373.948\\
  \end{array}
\right]
$
\end{tabular}
\end{table}

\begin{figure*}[!t]
\centering
\begin{tabular}{cc}
\subfloat[\label{subAnalysisA}Pareto plot of factor effects with respect to NPB Runtime.]{\includegraphics[width=3.3in]{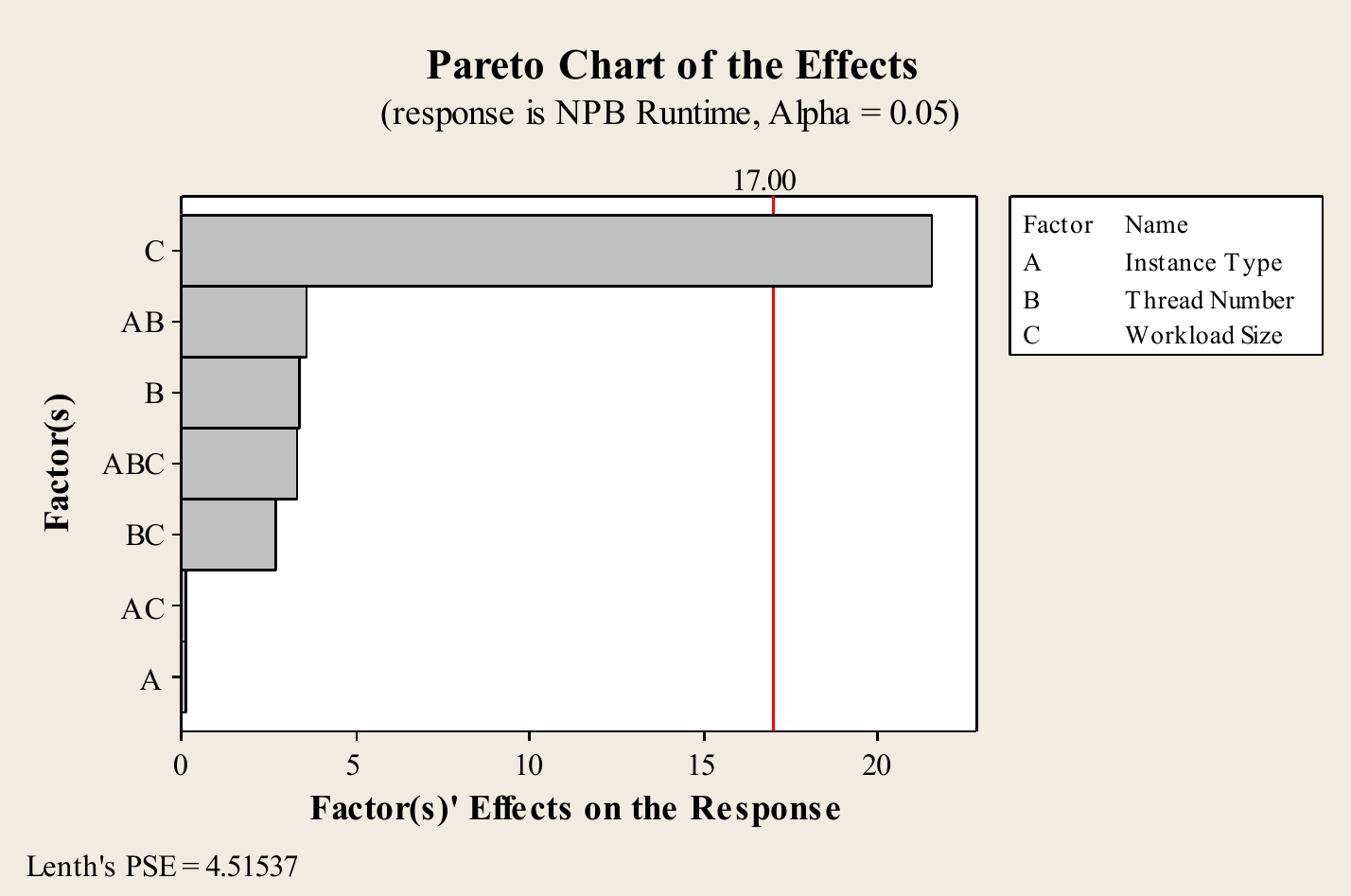}} & 
\subfloat[\label{subAnalysisB}Pareto plot of factor effects with respect to NPB FLOP Rate.]{\includegraphics[width=3.3in]{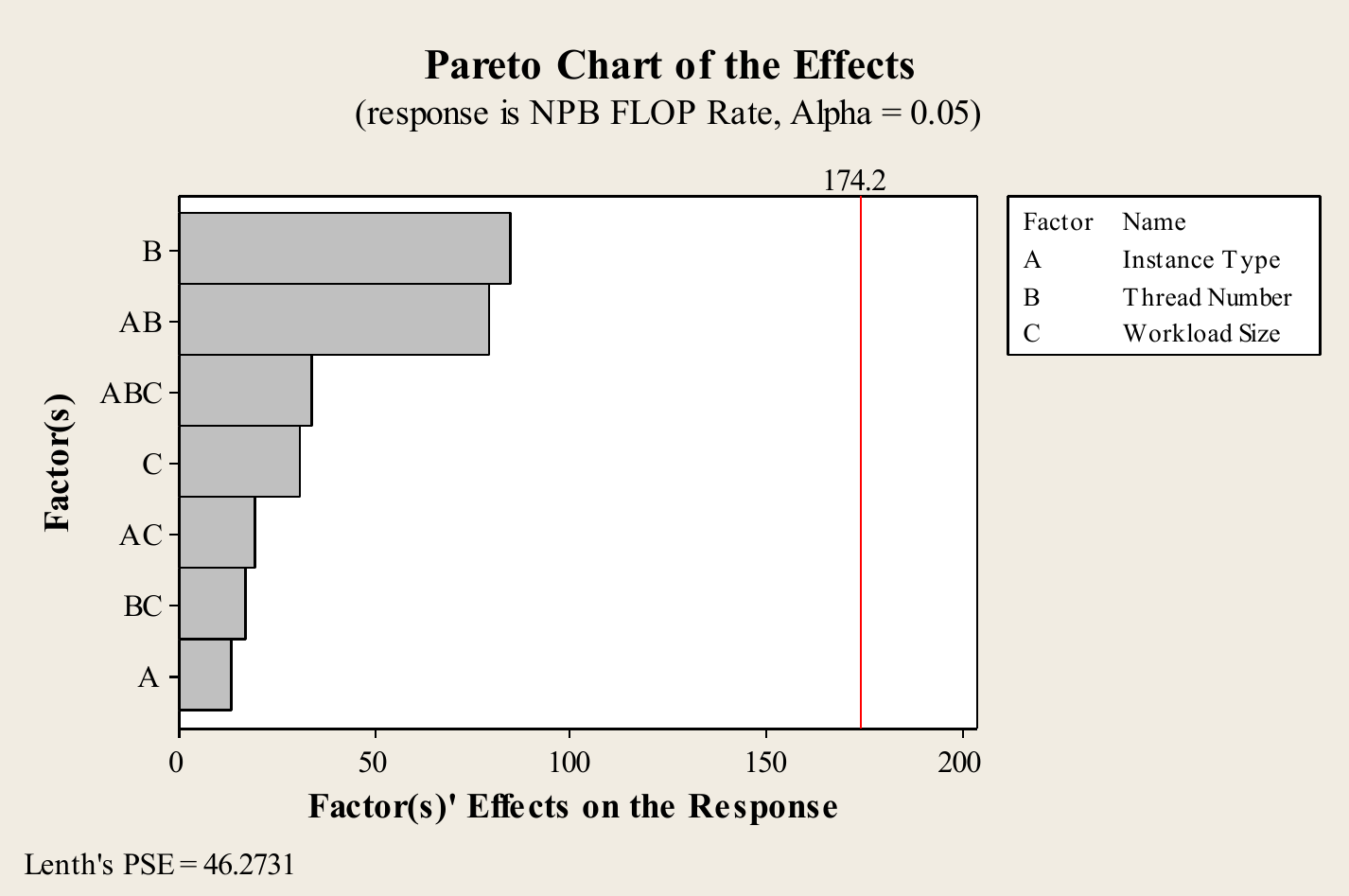}}\\
\end{tabular}
\caption{The Pareto plots of experimental factor effects (generated by Minitab).}
\label{figure>analysis}
\end{figure*}

Recall that the analysis is to investigate if \textit{Instance Type (A)} (or other factors) significantly influences the benchmarking results. By setting the significance level $\alpha$ as 0.05 \cite{Jackson_2011}, we can draw Pareto plots \cite{Antony_2003} to detect the factor and interaction effects that are important to the parallel computing (NPB suite in this case), as shown in Figure \ref{figure>analysis}. To save space, we do not elaborate the backend statistics here. In brief, given a particular significance level, Pareto plot displays a red reference line besides the effect values. Any effect that extends past the reference line is potentially important \cite{Antony_2003}.

Since the effect of factor \textit{Workload Size (C)} is beyond the reference line in Figure \ref{subAnalysisA}, it is apparent that \textit{Workload Size (C)} dominates the runtime of NPB suite. On the contrary, the factor \textit{Instance Type (A)} has little influence on the benchmark runtime in this case. As for the FLOP Rate analysis in Figure \ref{subAnalysisB}, we show that none of the factor or interaction effects significantly influences the transaction speed. However, relatively speaking, \textit{Thread Number (B)} is the most important to the NPB FLOP Rate, while \textit{Instance Type (A)} is still the least important factor. Therefore, for our proposed parallel computing project, we are now suggested to pay more attention to the workload size to distinguish between those two EC2 instance types. 

From Figure \ref{figure>result}, it is clear that increasing thread numbers will not increase computing performance if an instance's CPU cores are already saturated, especially with larger workload. 
However, increasing workload size seems to be able to continually increase the performance difference between two instances at 4 or more threads, which was further confirmed by running a supplementary experiment with NPB's 4X larger workload Class B. To facilitate analysis, we calculated different performance improvements of switching from m2.xlarge to m1.xlarge at 4 threads by using Equation (\ref{equation>improvement}), as listed in Table \ref{table>improvement}. Note that we use the minimum performance value between the two instances as the denominator to avoid the \textit{Ratio Game} bias \cite{Jain_1991}.

\begin{small}
\begin{equation}
\label{equation>improvement}
I=\frac{\left|Performance_{m2}-Performance_{m1}\right|}{MIN(Performance_{m1},Performance_{m2})}\times 100\%
\end{equation}
\end{small}

\begin{table}[!t]\footnotesize
\renewcommand{\arraystretch}{1.3}
\centering
\caption{\label{table>improvement}Performance Improvement of Switching from M2.xlarge to M1.xlarge at 4 Threads}
\begin{tabular}{|c|c|c|}
\hline
\textbf{\multirow{2}{*}{Workload}} & \multicolumn{2}{c|}{\textbf{Performance Improvement}}\\
\cline{2-3}
& \textbf{NPB Runtime} & \textbf{NPB FLOP Rate}\\
\hline
\textbf{Class W} & 9.4\% & 10.4\% \\
\hline
\textbf{Class A} & 39.6\% & 39.5\% \\
\hline
\textbf{Class B} & 48.4\% & 48.4\% \\
\hline
\end{tabular}
\end{table}

Given the price increase of switching from m2.xlarge to m1.xlarge ($(0.95-0.57)/0.57\times 100\%=61.4\%$), the instance-hour for running m2.xlarge is 1.614 times higher than running m1.xlarge within the same budget. In other words, m2.xlarge always has a cost advantage over m1.xlarge until the performance improvement reaches 61.4\%, although the total runtime may be longer. According to the previous analysis, we finally decided to choose m2.xlarge as the cost-wise option for our small-scale parallel computing project.


\section{Conclusions and Future Work}
\label{V}
Although benchmark suites have been identified as significant for, and been widely employed in, Cloud services evaluation, most evaluation studies intended to report individual benchmarking results without summarization. To help customers understand the holistic performance of a Cloud service, we suggest to adopt \textit{Boosting Metrics} to depict summary measures as single scores when using a suite of benchmarks to perform evaluation. In fact, delivering a single score is also a usual benchmarking strategy to facilitate drawing simple conclusions from evaluation results \cite{Islam_Lee_2012}, which can further facilitate applying experimental design and analysis to the evaluation work, as demonstrated in Section \ref{III}. Moreover, the idea of boosting metrics can be used to help measure a complex Cloud service feature involving multiple service properties. For example, although evaluating Elasticity of a Cloud service (covering both provisioning latency and cost) is not trivial \cite{Kossmann_Kraska_2010}, by integrating relevant basic QoS metrics to monitor the requested Cloud resources, our colleagues have developed a Penalty Model to measure the imperfections in elasticity of Cloud services for a given workload in monetary units \cite{Islam_Lee_2012}.

Our future work will be unfolded along two directions. Firstly, we plan to gradually collect, propose, and report new boosting metrics to supplement primary measures of individual Cloud service features. Secondly, we will concentrate on the Elasticity of Cloud services, and help improve the current approach \cite{Islam_Lee_2012} to Elasticity evaluation by employing suitable boosting metrics.









\begin{thebibliography}{99}
\itemsep 3pt

\bibitem{Akioka_Muraoka_2010}
S.~Akioka and Y.~Muraoka, ``HPC benchmarks on Amazon EC2," \emph{Proc.~WAINA 2010}, Apr.~2010, pp.~1029--1034.

\bibitem{Antony_2003}
J.~Antony, \emph{Design of Experiments for Engineers and Scientists}. Burlington, MA: Butterworth-Heinemann, Nov.~2003.

\bibitem{AWS_2012}
AWS, ``Amazon EC2 instance types," \emph{Amazon Web Services}, available at \url{http://aws.amazon.com/ec2/instance-types/}, Sept.~2012.

\bibitem{Buyya_Yeo_2009}
R.~Buyya, C.~S.~Yeo, S.~Venugopal, J.~Broberg, and I.~Brandic, ``Cloud Computing and emerging IT platforms: Vision, hype, and reality for delivering computing as the 5th utility," \emph{Future Gener.~Comp.~Sy.}, vol.~25, no.~6, Jun.~2009, pp.~599--616.

\bibitem{Brooks_2010}
C.~Brooks, ``Cloud Computing benchmarks on the rise," \emph{SearchCloudComputing, TechTarget}, available at \url{http://searchcloudcomputing.techtarget.com/news/1514547/Cloud-computing-benchmarks-on-the-rise}, Jun.~2010.

\bibitem{Cantrell_2012}
D.~W.~Cantrell, ``Pythagorean means," \emph{MathWorld--A Wolfram Web Resource}, available at \url{http://mathworld.wolfram.com/PythagoreanMeans.html}, Sept.~2012.

\bibitem{Carlyle_Harrell_2010}
A.~G.~Carlyle, S.~L.~Harrell, and P.~M.~Smith, ``Cost-effective HPC: The community or the Cloud," \emph{Proc.~CloudCom 2010}, IEEE Computer Society, Nov./Dec.~2010, pp.~169--176.

\bibitem{Dejun_Pierre_2009}
J.~Dejun, G.~Pierre, and C.-H.~Chi, ``EC2 Performance Analysis for Resource Provisioning of Service-Oriented Applications," \emph{Proc.~ICSOC/ServiceWave 2009}, Springer-Verlag, Nov.~2009, pp.~197--207.

\bibitem{Evangelinos_Hill_2008}
C.~Evangelinos and C.~N.~Hill, ``Cloud computing for parallel scientific HPC applications: Feasibility of running coupled atmosphere-ocean climate models on Amazon's EC2," \emph{Proc.CCA 2008}, Oct.~2008, pp.~1--6.

\bibitem{Ferdman_Adileh_2012}
M.~Ferdman, A.~Adileh, O.~Kocberber, S.~Volos, M.~Alisafaee, D.~Jevdjic, C.~Kaynak, A.~D.~Popescu, A.~Ailamaki, and B.~Falsafi, ``Clearing the Clouds: A study of emerging scale-out workloads on modern hardware," \emph{Proc.~ASPLOS 2012}, Mar.~2012, pp.~37--48.

\bibitem{He_Zhou_2010}
Q.~He, S.~Zhou, B.~Kobler, D.~Duffy, and T.~McGlynn, ``Case study for running HPC applications in public Clouds," \emph{Proc.~HPDC 2010}, ACM Press, Jun.~2010, pp.~395--401.

\bibitem{Iosup_Ostermann_2011}
A.~Iosup, S.~Ostermann, N.~Yigitbasi, R.~Prodan, T.~Fahringer, and D.H.J.~Epema, ``Performance analysis of Cloud computing services for many-tasks scientific computing," \emph{IEEE Trans.~Parallel Distrib.~Syst.}, vol.~22, no.~6, Jun.~2011, pp.~931--945.

\bibitem{Islam_Lee_2012}
S.~Islam, K.~Lee, A.~Fekete, and A.~Liu, ``How a consumer can measure elasticity for Cloud platforms," \emph{Proc.~ICPE 2012}, ACM Press, Apr.~2012, pp.~85--96.

\bibitem{Jackson_2011}
S.~L.~Jackson, \emph{Research Methods and Statistics: A Critical Thinking Approach}, 4th ed. Belmont, CA: Wadsworth Publishing, Mar.~2011.

\bibitem{Jackson_Ramakrishnan_2010}
K.~R.~Jackson, L.~Ramakrishnan, K.~Muriki, S.~Canon, S.~Cholia, J.~Shalf, H.~J.~Wasserman, and N.~J.~Wright, ``Performance analysis of high performance computing applications on the Amazon Web services Cloud," \emph{Proc.~CloudCom 2010}, IEEE Computer Society, Nov.-Dec.~2010, pp.~159--168.

\bibitem{Jain_1991}
R.~K.~Jain, \emph{The Art of Computer Systems Performance Analysis: Techniques for Experimental Design, Measurement, Simulation, and Modeling}. New York, NY: Wiley Computer Publishing, John Wiley \& Sons, Inc., May 1991.

\bibitem{Kossmann_Kraska_2010}
D.~Kossmann and T.~Kraska, ``Data management in the Cloud: Promises, state-of-the-art, and open questions," \emph{Datenbank Spektr.}, vol.~10, no.~3, Nov.~2010, pp.~121--129.

\bibitem{leBoudec_2010}
J.-Y.~Le Boudec, \emph{Performance Evaluation of Computer and Communication Systems}. Lausanne, Switzerland: EFPL Press, Oct.~2010.

\bibitem{Li_OBrien_2012b}
Z.~Li, L.~O'Brien, H.~Zhang, and R.~Cai, ``On a catalogue of metrics for evaluating commercial Cloud services," \emph{Proc.~Grid 2012}, IEEE Computer Society, Sept.~2012, pp.~164--173.

\bibitem{Li_OBrien_2012c}
Z.~Li, L.~O'Brien, H.~Zhang, and R.~Cai, ``A factor framework for experimental design for performance evaluation of commercial Cloud services," \emph{Proc.~CloudCom 2012}, IEEE Computer Society, Dec.~2012, to appear.

\bibitem{Li_Yang_2010}
A.~Li, X.~Yang, S.~Kandula, and M.~Zhang, ``CloudCmp: Comparing public Cloud providers," \emph{Proc.~IMC 2010}, ACM Press, Nov.~2010 pp.~1--14.

\bibitem{Li_Zhang}
Z.~Li, H.~Zhang, L.~O'Brien, R.~Cai, and S.~Flint, ``On evaluating commercial Cloud services: A systematic review," submitted to \emph{J.~Syst.~Software}.

\bibitem{Luszczek_Bailey_2006}
P.~R.~Luszczek, D.~H.~Bailey, J.~J.~Dongarra, J.~Kepner, R.~F.~Lucas, R.~Rabenseifner, and D.~Takahashi, ``The HPC Challenge (HPCC) benchmark suite," \emph{Proc.~SC 2006}, ACM Press, Nov.~2006, p.~213.


\bibitem{Montgomery_2009}
D.~C.~Montgomery, \emph{Design and Analysis of Experiments}, 7th ed. Hoboken, NJ: John Wiley \& Sons, Inc., Jan.~2009.

\bibitem{NASA_2012}
NASA, ``NAS Parallel Benchmarks," \emph{NASA Advanced Supercomputing Division}, available at \url{http://www.nas.nasa.gov/publications/npb.html}, Mar.~2012.

\bibitem{Obaidat_Boudriga_2010}
M.~S.~Obaidat and N.~A.~Boudriga, \emph{Fundamentals of Performance Evaluation of Computer and Telecommunications Systems}. Hoboken, New Jersey: Wiley-Interscience, Jan.~2010.

\bibitem{Ostermann_Iosup_2009}
S.~Ostermann, A.~Iosup, N.~Yigitbasi, R.~Prodan, T.~Fahringer, and D.~H.~J.~Epema, ``A performance analysis of EC2 Cloud computing services for scientific computing," \emph{Proc.~CloudComp 2009}, Springer-Verlag, Oct.~2009, pp.~115--131.

\bibitem{Prodan_Ostermann_2009}
R.~Prodan and S.~Ostermann, ``A survey and taxonomy of Infrastructure as a Service and Web hosting Cloud providers," \emph{Proc.~GRID 2009}, IEEE Computer Society, Oct.~2009, pp.~17--25.

\bibitem{Rabl_Frank_2010}
T.~Rabl, M.~Frank, H.~M.~Sergieh, and H.~Kosch, ``A data generator for Cloud-scale benchmarking," \emph{Proc.~TPCTC 2010}, ACM Press, Sept.~2010, pp.~41--56.

\bibitem{Schapire_2002}
R.~E.~Schapire, ``The boosting approach to machine learning: An overview," \emph{Proc.~MSRI Workshop on Nonlinear Estimation and Classification 2002}, Springer-Verlag, 2002, pp.~1--22.

\bibitem{Sobel_Subramanyam_2008}
W.~Sobel, S.~Subramanyam, A.~Sucharitakul, J.~Nguyen, H.~Wong, A.~Klepchukov, S.~Patil, A.~Fox, and D.~Patterson, ``Cloudstone: Multiplatform, multi-language benchmark and measurement tools for Web 2.0," \emph{Proc.~CCA 2008}, Oct.~2008, pp.~1--6.

\bibitem{Walker_2008}
E.~Walker, ``Benchmarking Amazon EC2 for high-performance scientific computing," \emph{;Login}, vol.~33, no.~5, Oct.~2008, pp.~18--23.


\end{thebibliography}
%

\end{document}